\documentclass[runningheads]{llncs}
\usepackage{graphicx}
\usepackage{amsmath}
\usepackage{booktabs}
\usepackage{times}
\usepackage{multirow}
\usepackage{amssymb}
\usepackage{xcolor}
\usepackage[misc]{ifsym}

\begin{document}
\title{Multi-level Adaptation of Distributed Decision-Making Agents in Complex Task Environments}
\titlerunning{Multi-level adaptation in complex tasks}
%
\author{Darío Blanco-Fernández \textsuperscript{(\Letter)}\orcidID{0000-0002-8275-701X} \and
Stephan Leitner\orcidID{0000-0001-6790-4651} \and
Alexandra Rausch\orcidID{0000-0002-9275-252X}}
\authorrunning{D. Blanco-Fernández et al.}
%
\institute{University of Klagenfurt, Klagenfurt, 9020, Austria\\ \email{\{dario.blanco, stephan.leitner, alexandra.rausch\}@aau.at}}
\maketitle              
\begin{abstract}
To solve complex tasks, individuals often autonomously organize in teams. Examples of complex tasks include disaster relief rescue operations or project development in consulting. The teams that work on such tasks are adaptive at multiple levels: First, by autonomously choosing the individuals that jointly perform a specific task, the team itself adapts to the complex task at hand, whereby the composition of teams might change over time. We refer to this process as self-organization. Second, the members of a team adapt to the complex task environment by learning.
There is, however, a lack of extensive research on multi-level adaptation processes that consider self-organization and individual learning as simultaneous processes in the field of management science. We introduce an agent-based model based on the \textit{NK}-framework to study the effects of simultaneous multi-level adaptation on a team's performance. 
We implement the multi-level adaptation process by a second-price auction mechanism for self-organization at the team level. Adaptation at the individual level follows an autonomous learning mechanism. 
Our preliminary results suggest that, depending on the task's complexity, different configurations of individual and collective adaptation can be associated with higher overall task performance. Low complex tasks favour high individual and collective adaptation, while moderate individual and collective adaptation is associated with better performance in case of moderately complex tasks. For highly complex tasks, the results suggest that collective adaptation is harmful to performance.

\end{abstract}
\section{Introduction}
Disaster relief rescue operations \cite{Tang2018}, or project development in consulting firms \cite{Creplet2001} are examples of tasks that can be characterized as \textit{complex tasks}. 
These and many other complex tasks have two characteristics in common: (i) A single individual alone usually cannot find a solution to complex tasks, as the capabilities required to solve them are greater in scope than the ones a single individual possesses \cite{Funke1995}, and (ii) complex tasks are formed by various subtasks that are interdependent \cite{Giannoccaro2019}.

Individuals are required to coordinate and share their capabilities to solve complex tasks. We refer to this process as \textit{team} formation \cite{Simon1957}. Team formation often occurs autonomously, i.e., agents form teams by themselves and without the direct intervention of a central planner or decision-maker \cite{Licalzi2012}. Giving individuals the possibility to autonomously organize themselves into teams can, for example, be found in firms that engage in consulting work \cite{Creplet2001}, or professional services firms that are organized as a partnership (such as law or accounting firms) \cite{Gershkov2009}. 

Teams that engage in complex problem solving might be dynamic in their composition, as they change by adapting to the particular task they face \cite{Creplet2001}. Adaptation at the team level is referred to as \textit{collective adaptation}. However, not only the team as a collective adapts but also the individual agents (who make up a team) adapt by learning \cite{Creplet2001} (i.e.,  agents go through a process of \textit{individual adaptation}). Thus, a team that is formed to solve a complex task goes through a continuous \textit{multi-level} adaptation process. By successfully adapting at the individual and collective level to a particular complex task, the team's performance is expected to improve \cite{Eisenhardt2000}.

Previous research indicates that individual agents adapt their capabilities to the task requirements by learning about the task they face \cite{Baumann2019}, 
becoming more capable of performing the task, and thus improving the overall \textit{task performance} \cite{Eisenhardt2000}. To model individual adaptation, we consider a learning approach that can be characterized as \textit{autonomous}. Learning is autonomous when it occurs without any interaction between agents, or a central agent who can take the role of a supervisor or a teacher. Similar autonomous approaches to learning can be found in \cite{leitner14,Levinthal1997}. 

Individual agents are endowed with the ability to adapt and to autonomously form teams. Research in economics and management science often assumes that agents are homogeneous (i.e., following the concept of the \textit{representative agent}) \cite{Axtell2007}. This assumption implies that any collective (e.g., a team) is just the aggregation of a set of homogeneous agents; and allows to study collective adaptation similarly to individual learning, since teams can be treated as uniform entities that go through a process of learning \cite{Rivkin2007}. 
However, the notion of homogeneity in agents poses a problem, as this assumption is rather unrealistic and does not reflect real-life settings properly: Individuals usually differ in their characteristics \cite{Axtell2007}. In contrast, by considering heterogeneous agents, we drop this restrictive assumption and give this research a more realistic perspective \cite{Wall2016}. 
Moreover, it also allows for implementing collective adaptation as a process differing from just the aggregation of individual learning dynamics. For example, by autonomously forming and recurrently reorganizing teams, a continuous process of collective adaptation emerges by reconsidering the role of the current members of the team against other potential members with different capabilities \cite{Gomes2019}. A significant branch of the previous literature concerning team self-organization has implemented auction-based mechanisms for team formation, in which an auction is held and the highest bidders are selected to form the team \cite{Rizk2019}. 
We consider collective adaptation as a self-organization process which follows an auction-based approach.

It seems reasonable to think that more adaptation is always positive for performance. However, in contrast to popular belief, more knowledge can be a burden. Previous research has shown that individual learning is beneficial for task performance, but only up to a threshold \cite{Levinthal1997,March1991}. Moreover, researchers argue that recurrent team self-organization can be harmful to performance in certain circumstances \cite{Gomes2019}. This is related to the \textit{exploration vs exploitation dilemma}, which states that individuals should adequately balance searching for new solutions (i.e., exploration) against building on the solutions already at their disposal (i.e., exploitation) \cite{March1991}. According to \cite{Levinthal1997}, the emergence of new solutions is positive for performance only at early stages of task-solving. However, at later stages, acquiring new solutions instead of building on the current ones known can lead to sub-optimal situations in terms of task performance \cite{Levinthal1997}. We consider a multi-level perspective on this dilemma: Exploration can be increased (decreased) either by increasing individual learning or by self-organizing the team more often. Both actions imply that the team is actively looking for new solutions. Decreasing individual learning or self-organization, in turn, can be associated with more exploitation of the current solutions available.


Given the vast range of complex tasks in real-life settings \cite{Creplet2001,Hsu2016,Tang2018}, it is important to understand the relationship between simultaneous adaptation at multiple levels and complex task performance. While research on multi-level adaptation for task solving can be found in fields such as physics \cite{Gomes2019}, it has not been extensively studied in the field of management science. This is done due to the fact that the micro and the macro-level (i.e., the individual and the team level) of the studied systems have been traditionally considered separately, and the literature that unifies both into a single framework is relatively scarce \cite{Molloy2011}. This aspect points at a significant research gap that we aim to fill. By implementing a multi-level adaptation approach to a practical management science problem, our objective is to understand better how simultaneous individual and collective adaptation affect task performance. To do so, we propose an agent-based model based on the \textit{NK}-framework for management science \cite{Levinthal1997} and perform a simulation study. 
This paper is structured as follows: Sec. \ref{sec:model} discusses the model and its most important aspects. Results are presented in Sec. \ref{sec:results}. Finally, Sec. \ref{sec:summary} provides a discussion of the results and potential avenues for future research.
\section{Model}\label{sec:model}
\begin{figure}
      \centering
      \includegraphics[width=0.8\linewidth]{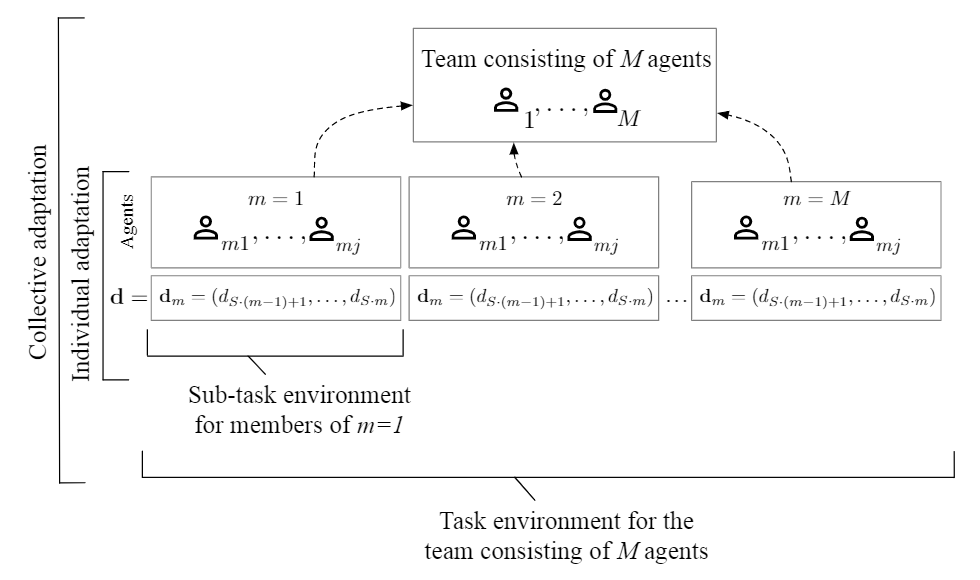}
      \caption{Model architecture}
      \label{fig:structure}
  \end{figure}

We model situations in which a population of $P\in \mathbb{N}$ agents faces a task of $N\in\mathbb{N}$ binary choices (see Sec. \ref{sec:task-environment}). The agents' capabilities are limited in the following two ways: (i) Single agents cannot solve the task on their own, so they have to collaborate with other agents in a team consisting of $M\in\mathbb{N}<P$ members to jointly find a solution to the task; and (ii) agents cannot evaluate the entire set of possible solutions simultaneously. To overcome limitation (i), agents are endowed with the capability of recurrently self-organizing into a team (see Sec. \ref{sec:collective-adaptation}). Concerning limitation (ii), we endow the agents with the capability of adapting to the task environment by sequentially exploring the solution space and learning new solutions in the process. We refer to this process as individual adaptation (see Sec. \ref{sec:individual-adaptation}). The model architecture is illustrated in Fig. \ref{fig:structure}.  We observe for $t=\{1,\dots,T \}\subset\mathbb{N}$ periods, how individual and collective adaptation affect task performance and interact. 

\begin{figure}
      \centering
      \includegraphics[width=\linewidth]{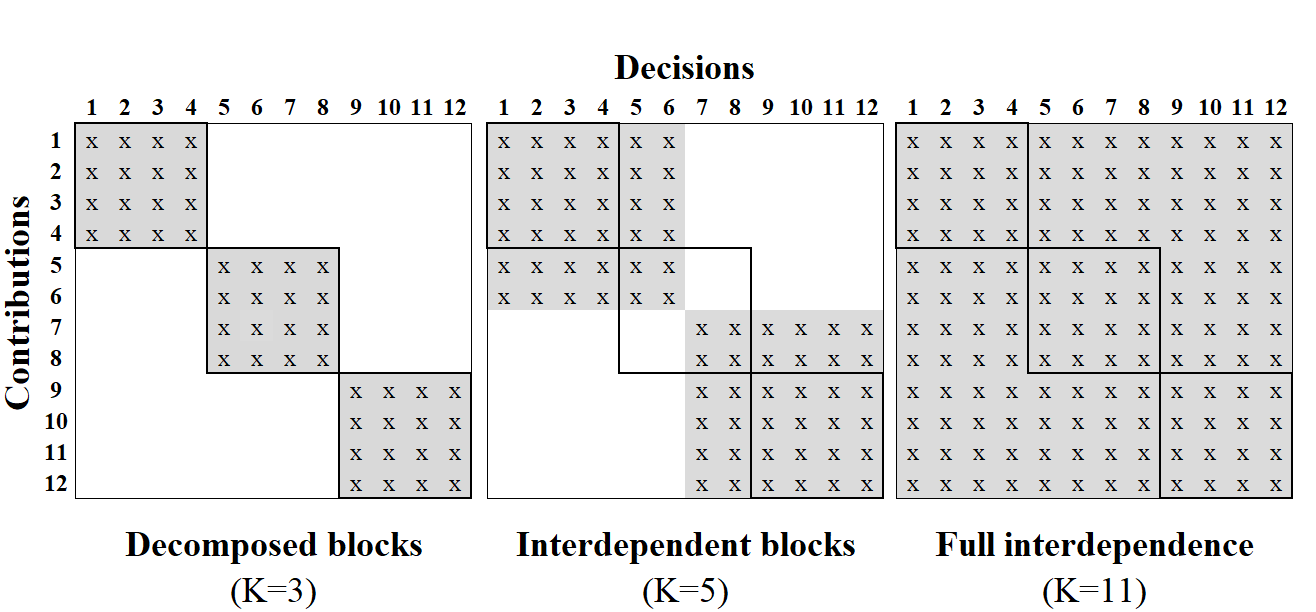}
      \caption{Stylized interdependence structures}
      \label{fig:interdependencies}
  \end{figure}

\paragraph{Task environment and agents}
\label{sec:task-environment}
We base the task environment on the \textit{NK}-framework \cite{Levinthal1997} and formalize the complex decision problem, which agents face, by the string $\mathbf{d}=(d_{1}, \dots,d_{N})$, where $d_n\in \{0,1\}$ for $n=\{1,\dots,N\}\subset\mathbb{N}$. Since agents are limited in their capabilities (see Sec. \ref{sec:individual-adaptation}), we segment the decision problem into $M$ parts of equal size $S\in\mathbb{N}=N/M$. Within the team, agent $m=\{1,\dots,M\}\subset\mathbb{N}$ is responsible for the subtasks $\mathbf{d}_{m} = (d_{S\cdot(m-1)+1},\dots,d_{S\cdot m})$.
Each decision $d_n$ is associated with a performance contribution $f(d_n) \sim U(0,1)$. $K \in \mathbb{N}_0$ interdependencies among decisions $d_i$ shape the complexity of the task $\mathbf{d}$, so that performance contribution $f(d_n)$ might not only be affected by $d_n$ but also by $K$ other decisions. $K$ represents the \textit{complexity} of the overall task. We formalize the corresponding pay-off function by 
\begin{equation}
f(d_n)=f(d_n, \underbrace{d_{i_1}, \dots, d_{i_K})}_{K \textrm{interdependencies}}~, 
\label{eq:payoff}
\end{equation}
\noindent where $\{i_1, \dots, i_K\}\subseteq \{1, \dots, ,n-1, n+1,\dots, N\}$, and $0\leq K \leq N-1$. To compute performance landscapes\footnote{For a detailed discussion of the resulting landscapes' characteristics, the reader might consult \cite{Levinthal1997}} of different complexity based on Eq. \ref{eq:payoff}, we consider the stylized interdependence structures illustrated in Fig. \ref{fig:interdependencies}.

We denote agent $m$'s solution to their problem $\mathbf{d}_m$ and the solution to the entire problem $\mathbf{d}$ at time step $t$ by $\mathbf{d}_{mt}$ and $\mathbf{d}_t$, respectively. We compute agent $m$'s performance in $t$ according to 
\begin{equation}
    \phi(\mathbf{d}_{mt}) = \frac{1}{S} \sum_{d_{nt} \in \mathbf{d}_{mt}} f(d_{nt})~ 
\end{equation}
\noindent and the performance associated with the solution of the the team of $M$ agents by 
\begin{equation}
    \Phi(\mathbf{d}_t) = \frac{1}{M} \sum_{m=1}^{M} \phi (\mathbf{d}_{mt}) = \frac{1}{N} \sum_{n=1}^{N} f(d_{nt})~. 
    \label{eq:performance}
\end{equation}

\noindent Agent $m$'s utility in time step $t$ includes the performance of their subtasks $\mathbf{d}_m$ and the residual performance coming from the other $M-1$ agents' decisions $\mathbf{D}_{m}=(\mathbf{d}_1,\dots,\mathbf{d}_r)$, where $r=\{1,\dots,M\}$ and $r\neq m$. Agent $m$'s utility in $t$ follows the linear function

\begin{equation}
   U(\mathbf{d}_{mt}, \underbrace{\mathbf{D}_{mt}}_{(\mathbf{d}_{1t},\dots,\mathbf{d}_{rt})}) = \alpha \cdot \phi(\mathbf{d}_{mt}) + \beta \cdot \frac{1}{M-1} \sum_{\substack{r=1 \\ r\neq m}}^{M}\phi(\mathbf{d}_{rt}) ~,
   \label{eq:utility}
\end{equation}

\noindent where $\alpha \in \mathbb{R}$ and $\beta \in \mathbb{R}$ indicate the weights for agent $m$'s own and residual performances, respectively, and $\alpha+\beta =1$. The objective of the agents is to maximize the utility function $U(\mathbf{d}_{mt}, \mathbf{D}_{mt})$ at each time step $t$.

\paragraph{Adaptation at the individual level}
\label{sec:individual-adaptation}
Since agents face binary choices, there are $S^2$ possible solutions for each subtask. Agents are limited as they do not know the entire solution space within their subtask. Initially, they only know a subset of $Q \in \mathbb{N} < S^2$ solutions. Agents are homogeneous concerning the capacity $Q$ and heterogeneous with respect to the $Q$ solutions they actually know. We endow agents with the capability of adapting to their subtask environment that is defined by their decision problem $\mathbf{d}_m$ (see Fig. \ref{fig:structure}). With probability $p$, agents randomly learn an unknown solution to $\mathbf{d}_m$ that differs in the value of only one decision ${d}_n$ from the set of currently known solutions. Since capabilities are limited, with probability $p$, agents also erase from their memory (i.e., forget) solutions they do not frequently use due to the low level of associated utility. The implemented adaptation follows an autonomous learning approach that can be found in previous implementations of the \textit{NK}-framework \cite{leitner14}.

\paragraph{Adaptation at the collective level}
\label{sec:collective-adaptation}
We assume that the population of $P$ agents is equally distributed across $M$ subtasks so that for each subtask, there are $J\in \mathbb{N} = P/M$ agents who can find a solution to this subtask. Consequently, $J$ agents compete for a slot in the team; and all $P$ agents autonomously organize themselves in a team finally composed of $M$ agents. The self-organizing process follows the concept of a second-price auction and always occurs at timestep $t=1$. Afterwards, $\tau-1$ other auctions are held over the remaining $T-1$ time steps in regular intervals, so for $t>1$, auctions are held each time $t\mod\frac{T}{\tau}=0$. This means that $\tau$ auctions take place over the considered time horizon $T$ each $\frac{T}{\tau}$ time steps.

We implement this second-price auction process in the following way: To become a team member, agents place bids. These bids represent what agents intend to contribute to team performance. Since the selection mechanism follows a second-price auction, agents have incentives to reveal their true contributions \cite{Vickrey1961}. Agents, however, cannot observe the other agents' bids and this is why they assume the residual decisions $\mathbf{D}_{mt-1}$ will remain constant from implemented solution at $t-1$. Agents compute the expected utilities for all their solutions to their subtasks $\mathbf{d}_m$ available in time step $t$ according to $U(\mathbf{d}_{mt},\mathbf{D}_{mt-1})$ (see Eq. \ref{eq:utility}). Each agent submits the highest attainable expected utility among the different solutions as their bid in $t$. 
Consequently, there are $J$ bids for each slot in the team. Following the logic of second-price auctions, the team comprises those agents $M$ who submit the highest proposals per slot. For being a part of the team, agents are charged the second-highest bid per slot.\footnote{Since agents only experience utility if they are team members, they always have incentives to participate in this process.}  

\paragraph{Decision-making process}
\label{sec:team-decision}
In each time step $t$, the $M$ team members are tasked with finding a solution $\mathbf{d_t}$ to the complex task $\mathbf{d}$. Agents are autonomous in their decisions, and there is no communication or coordination between agents. The decision-making process is a two-stage process. First, each agent $m$ focuses on their subtask $\mathbf{d}_{mt}$ and computes the expected utility for each solution available in this particular time step. Since we do not allow for communication, the agents assume that the residual decisions do not change compared to the previous period $\mathbf{D}_{mt-1}$. The expected utility is computed by $U(\mathbf{d}_{mt}, \mathbf{D}_{mt-1})$ (see Eq. \ref{eq:utility}) for each solution known. The agent chooses the solution that promises the highest utility. In the first step, each agent comes up with an $S$ binary values vector that represents the solution to their subtask.
Second, once all agents have submitted their solutions, the overall solution is computed by concatenating each of the $M$ solutions chosen by the agents. The solution to the complex problem in $t$, $\mathbf{d}_t,$ is thus represented by a vector of $N$ binary values. We compute the performance of the overall solution according to Eq. \ref{eq:performance}, and agents receive the resulting utility according to Eq. \ref{eq:utility}.

\paragraph{Scenarios, variables and sequencing of the model}\label{sec:scenarios}
An overview of the model's variables and the values they can take is given in Tab. \ref{tab:variable}. With the values provided in Tab. \ref{tab:variable} for the number of auctions that occur at each simulation, we identify three different scenarios. First, for $\tau=1$, there is \textit{initial} collective adaptation: The auction process occurs only at the first period, never to be repeated. Second, for $\tau=20$, there is \textit{moderate} collective adaptation, with the team self-organizing each $\frac{T}{\tau}=10$ time steps. Finally, for $\tau=200$, there is \textit{high} collective adaptation, with the team self-organizing at each time step.

\begin{table*}[t]
\caption{Variables of the model}
\label{tab:variable}
\resizebox{\textwidth}{!}{%
\begin{tabular}{llcc}
Type                                & Description                          & \multicolumn{1}{c}{Denoted by}       & \multicolumn{1}{c}{Values} \\ \hline
\multirow{3}{*}{Exogenous variable} & Complexity                           & $K$                                  & $\{3, 5, 11\}$         \\
                                    & Probability of individual adaptation & $p$                                  & $\{0, 0.1, 0.2, 0.3, 0.4, 0.5\}$           \\
                                    & Number of auctions held          & $\tau$                               & $\{1, 20, 200\}$           \\ \hline
Observed variable                   & Team performance                    & $C_t$                                & $\in[0, … , 1]$           \\ \hline
\multirow{4}{*}{Other variables}   & Time steps                           & $t$                                  & $\in[0, … , 200]$           \\
                                    & Temporal horizon                     & $T$                                  & $200$           \\
                                    & Number of decisions                  & $N$                                  & $12$           \\
                                    & Weights of utility                   & $\alpha$, $\beta$                    & $0.5$           \\ & Simulation runs                   & $R$                    & 1,500            \\ & Number of subtasks                  & $M$                    & $3$ \\ \hline
\end{tabular}
}
\end{table*}

\begin{figure}[htp]
      \centering
      \includegraphics[width=\linewidth]{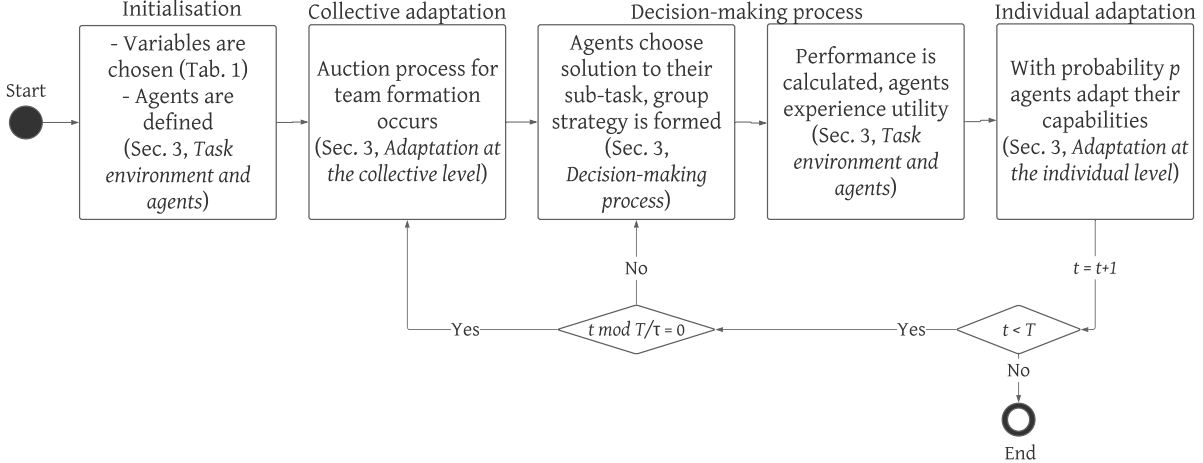}
      \caption{Sequence of events during simulation runs}
      \label{fig:sequence}
  \end{figure}

Fig. \ref{fig:sequence} provides an overview of the events of the model in order.

\section{Results}\label{sec:results}
\paragraph{Performance measure}
During each simulation run $r$ and timestep $t$, the team performance $\Phi(\mathbf{d}_t)$ is normalized by the highest possible performance in that simulation run, $max(\Phi_r)$. Based on this measure, we compute the average performance of the scenario for each timestep $t$ as follows:

\begin{equation}\label{eq:norm}
\tilde{\Phi_t}=\frac{1}{R}\sum_{t=1}^{T}\frac{\Phi(\mathbf{d_t})}{max(\Phi_r)} ~.   
\end{equation}

\noindent We further condense the average distance per timestep, $\tilde{\Phi_t}$, to the total Manhattan Distance. This means that we compute the difference at each time step $t$ between average performance $\tilde{\Phi_t}$ and the maximum performance $max(\Phi_r)$. Then each difference is summed over the $T$ time steps\footnote{Note that since performance has been normalized using Eq. \ref{eq:norm} and then averaged, then the maximum performance is $max(\Phi_r)=1$.}. We formalize the total Manhattan Distance \textit{MD} as follows:

\begin{equation}\label{eq:totdist}
    \textit{MD} = \sum^{T}_{t=1}{(1-\tilde{\Phi_t})}
\end{equation}

\noindent Please note that a higher Manhattan Distance \textit{MD} implies a higher average distance to the maximum, and thus a lower overall performance.

\begin{figure*}
      \includegraphics[width=\linewidth]{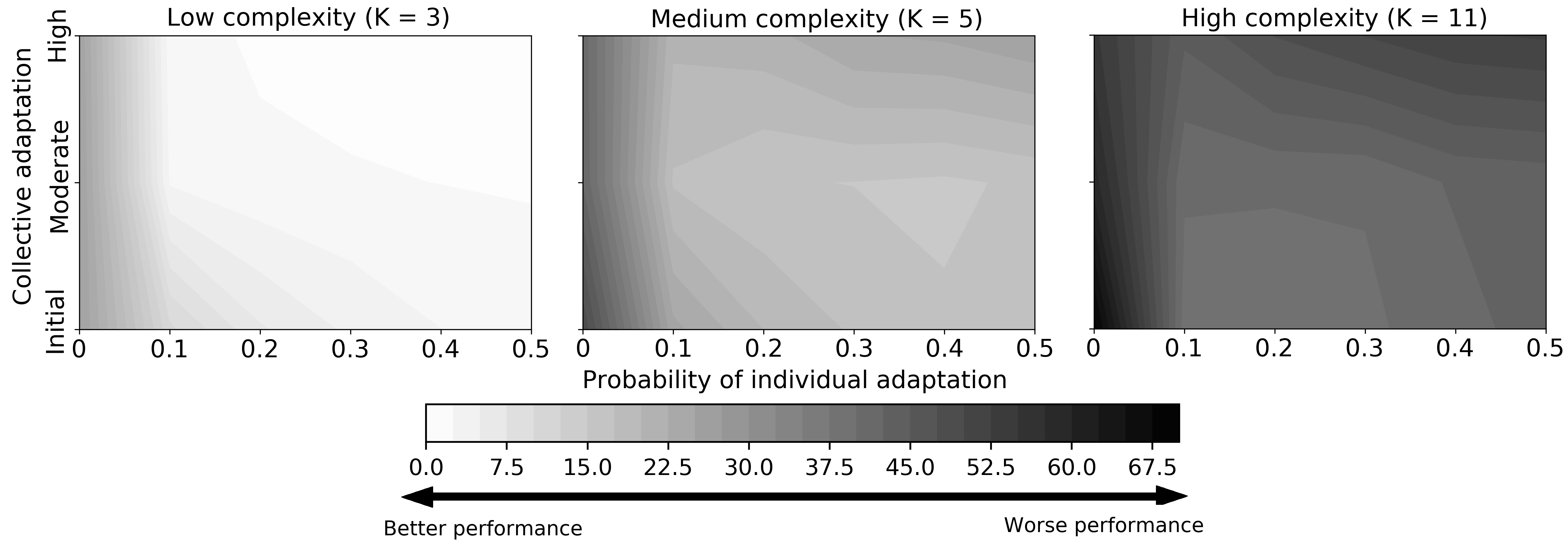}
      \caption{Contour plots for the total Manhattan Distance \textit{MD}.}
      \label{fig:results}
  \end{figure*}

In Fig. \ref{fig:results}, we plot the Manhattan Distance for each studied scenario. The x-axis of each subplot represents the probability of individual adaptation. On the y-axis, we plot the three scenarios considered for collective adaptation (see Sec. \ref{sec:scenarios}, \textit{Scenarios, variables and sequencing of the model}). The three subplots represent the results for different complexity levels (see Fig. \ref{fig:structure}). Each point of the subplots represents a total Manhattan Distance \textit{MD} for a particular configuration of individual and collective adaptation. A point located in a lighter (darker) area implies lower (higher) \textit{MD}, and thus a better (worse) performance.

\paragraph{Individual adaptation} The results suggest that the probability of individual adaptation $p$ has a considerable effect on task performance. In particular, we observe a substantial decrease in the total Manhattan Distance \textit{MD}, when agents are capable of learning, i.e., when $p>0$. 
This pattern is robust across all considered interdependence structures (i.e., for each level of $K$ considered), as shown in all three contour plots of Fig. \ref{fig:results}. The effect of agents being capable of learning on performance is highly significant, suggesting that promoting learning, even if it is just minimal, is always beneficial for task performance.

Regarding $p>0.1$, it can be observed in low complexity scenarios (i.e., the left plot of Fig. \ref{fig:results}) that there is a decreasing marginal positive effect of individual adaptation on performance: Each successive increase in $p$ increases task performance less than the previous increase did. 
For example, 
the effect on the average task performance of increasing the probability of individual adaptation from $p=0.1$ to $p=0.5$ is lower in magnitude than the effect of increasing the probability from $p=0$ to $p=0.1$ for all three collective adaptation scenarios considered. Starting to learn about a task has huge implications on performance. However, each increase that occurs in individual adaptation reduces the overall impact of individual learning on performance, even if this effect is still positive.

However, for higher levels of complexity (see the central and right panels of Fig. \ref{fig:results}), this positive effect eventually turns negative, and the overall performance declines as a result. For example, in the central plot of Fig. \ref{fig:results} (i.e., for $K=5$) we observe how a team with high collective adaptation reaches its lowest total distance of $MD=20.00$ at a learning probability of $p=0.1$. Further increases in individual adaptation appear to affect performance negatively, surpassing $MD=25.00$ for $p=0.5$. Eventual decreases in performance can also be observed for moderate collective adaptation and for 
initial collective adaptation (both for $p>0.4$).
This pattern of initial growth but with an eventual decline in performance is even more pronounced in scenarios of high complexity (i.e., $K=11$, see the right panel of Fig. \ref{fig:results}). 
Results imply that, for moderate or high levels of complexity, individual adaptation and task performance are related in the form of an inverted-U: Initial increases in individual adaptation are positive for task performance, although the positive effect decreases in magnitude with each successive increase. Eventually, the relationship turns negative and performance is harmed as a consequence of increasing individual adaptation.

\paragraph{Collective adaptation} The results suggest that whether collective adaptation has a positive or negative impact on performance depends on the complexity of the scenario studied. For low levels of complexity (represented by the left panel of Fig. \ref{fig:results}), we can observe that high collective adaptation is associated with higher performance as compared to scenarios with either moderate or initial collective adaptation (except for $p=0$, in which all three teams perform very similarly, and $p=0.1$, in which high and moderate collective adaptation perform similarly). 
Moreover, the highest possible performance in low complexity scenarios considered is attained by a team exhibiting high collective adaptation, reaching a total distance below $MD=1.00$ for $p=0.5$. When facing a task that is not highly complex, it is beneficial in terms of performance to increase the effort dedicated to finding new solutions (i.e., increasing exploration), at both the individual and collective level.

However, in scenarios with medium complexity (represented by the central panel of Fig. \ref{fig:results}), for most values of $p$, moderate collective adaptation leads to higher performances compared to both high or initial collective adaptation scenarios. There are two exceptions to this, which are for the extreme values of $p=0$ (in which moderate and high collective adaptation perform very similarly) and $p=0.5$ (in which moderate and initial collective adaptation perform very similarly). For $0<p<0.5$, moderate collective adaptation is associated with a higher performance than any other alternative. Moreover, the highest attainable performance in all medium complexity scenarios considered occurs for moderate collective adaptation and $p=0.4$ and is around $MD=15.00$. In contrast to low complexity scenarios (except for the extreme case of $p=0$, in which agents do not learn at all), high collective adaptation is always associated with the lowest performance compared to the other two alternatives. The results suggest that increasing exploitation (exploration) excessively via decreasing (increasing) collective adaptation can be harmful to performance. Moderate collective adaptation, in turn, benefits performance in the majority of cases.

Results in scenarios of high complexity, as shown in the right panel of Fig. \ref{fig:results}, suggest that increasing collective adaptation is detrimental to performance when individual learning occurs (i.e., for $p>0$).
An exception to this occurs in the extreme case of $p=0.5$, in which initial and moderate collective adaptation are associated with a similar performance around $MD=43.00$. For $0<p<0.5$, initial collective adaptation is associated with a higher performance than any other of the alternatives. Furthermore, the highest attainable performance for high complexity scenarios is attained by a team with initial collective adaptation and $p=0.2$, for approximately $MD=38.00$. When tasks are highly complex, stability in team composition is favoured, since the recurrent reorganization of a team eventually impairs task performance.

\paragraph{Discussion} 
This research is concerned with multi-level adaptation and its relationship with task performance in teams. Results indicate that, in low complexity environments, putting a lot of effort into adaptation at both the collective and the individual level is not detrimental in terms of performance. This extends the insights given by \cite{Levinthal1997} about adaptation in complex environments: When facing a task of low complexity, high exploration does not decrease the overall task performance, since there are lower chances of ending in sub-optimal, long-term solutions. Since only a single level at which adaptation occurs was considered in \cite{Levinthal1997}, our results contribute to previous research by showing that this is the case also for multi-level settings.

For moderate levels of complexity, our results show that the exploration vs exploitation dilemma can also be applied in a multi-level adaptation setting; and that a proper balance between exploration and exploitation is the key to improving task performance in teams. Thus, these insights are in line with previous research \cite{Levinthal1997,March1991,Rivkin2007,Wall2018} regarding the exploration vs exploitation dilemma. The main difference to the previous literature is that, in our research, exploration can be increased either by increasing individual or collective adaptation. In particular, when considering moderately complex tasks, the results suggest that for performance to improve, moderate individual learning has to be combined with moderate collective adaptation. The reported findings are consistent with previous results, which indicate that recurrent self-organization of teams combined with individual learning can be associated with higher task performance \cite{Gomes2019}. In real-life settings this moderate adaptation at both levels could be achieved, for example, by allowing employees of a firm to dedicate a limited amount of working hours to self-education and training; combined with holding regular team meetings with employees outside the team to discuss the state of the task and what other members could contribute to it.

Given these results for low and moderate complexity, it can be concluded that recurrent self-organization can be beneficial to task performance. These insights are in contrast with views in the field of management science that characterize team reorganization as harmful to performance \cite{Hsu2016}. In particular, team stability has been associated with high degrees of team-level learning and, eventually, higher performance \cite{Hsu2016}. In this context, adaptation occurs just at the team level, following a learning process for the collective. This highlights the importance of considering multi-level adaptation, as done in our research. By considering multiple, simultaneous adaptation processes (in this case, by endowing agents with the capabilities of learning and combined with recurrent team self-organization), emerging insights into topics such as team reorganization differ from those found in the previous literature. Since adaptation has been considered as occurring at multiple levels in the previous literature \cite{March1991}, we believe that these findings and similar implementations help us to understand better the relationship between adaptation and task solving in teams and to relate it to real-life problems.

Our results also contribute to the literature on multi-level adaptation and task performance when considering highly complex tasks. Previous research on multi-level adaptation and task performance (which is, in itself, not extensive) has not considered highly complex tasks (see, for example, \cite{Gomes2019}). Thus, we can consider our results as a first step in the matter of studying multi-level adaptation under high complexity. We find that, in highly complex tasks, collective adaptation is negative for performance, and that the positive effect of individual adaptation is limited considerably. As the previous literature shows, highly complex environments are associated with lower benefits for exploration, since the probability of finding sub-optimal, long-term solutions is very high \cite{Levinthal1997}. Again, our results can be understood as extensions to previous insights on the exploration vs exploitation dilemma by considering multi-level adaptation. 


This set of results provides an extension to approaches that consider a single level of adaptation \cite{Levinthal1997,March1991,Rivkin2007,Wall2018}. The value of our research lies in the fact that it provides a discussion on individual and collective adaptation, both separately and  simultaneously. To our knowledge, there is a lack of an extensive implementation of a multi-level adaptation approach that considers autonomous individual learning and team self-organization in management science. By considering these two aspects, we believe that this research provides a novel approach to multi-level adaptation in this field. Our results should then be understood as both complementary to previous findings in the literature and as an extension to them \cite{Levinthal1997,Wall2018}. 
\section{Conclusion}\label{sec:summary}
Individual and collective adaptation are two factors that strongly determine the performance of a task. In particular, we have found that regarding low complexity tasks, pushing for high individual and collective adaptation is associated with a better task performance. However, this is changed by increasing complexity. When tasks are moderately complex, teams have to combine individual learning up to a threshold with moderately recurrent team self-organization to reach the highest attainable performance. If the task is highly complex, then low individual learning combined with no collective adaptation is the best alternative in terms of task performance. These aspects provide evidence for the exploration vs exploitation dilemma also holds true when considering multi-level adaptation.

However, we must note that there are some limitations to our research. For example, the importance of coordination and communication is not addressed \cite{Wall2018}, and neither we do consider the potential costs that individual and collective adaptation might incur. Further extensions of the model include endowing agents with the capabilities of choosing more or less individual adaptation \cite{Giannoccaro2019} or breaking up the formed team \cite{Gomes2019}, and including alternative mechanisms of adaptation such as social learning \cite{Baumann2019}. The implementation of these and others aspects might help in extending our research. Moreover, the results suggest the existence of an interaction effect between adaptation at multiple levels, as increasing adaptation at both levels simultaneously can eventually lead to having "too much exploration" and decreasing task performance. Further extensions of this research can cover this aspect. Considering these limitations and potential extensions, we believe this paper and its results can serve as a first step towards the study of multi-level adaptation and its effects on performance. Moreover, it also serves as a departure point for future research on the topic of multi-level adaptation and complex task performance in teams.
%
%
%
\bibliographystyle{splncs04}
\bibliography{mybibliography}

\end{document}